\documentclass[%
reprint,
superscriptaddress,
amsmath,amssymb,
aps,
]{revtex4-2}

\usepackage{graphicx}
\usepackage{dcolumn}
\usepackage{bm}
\usepackage{hyperref}
\usepackage{csquotes}
\usepackage[
margin=0.6in,
]{geometry}

\begin{document}

\preprint{APS/123-QED}

\title{Scattering-robust Imaging of Azimuthal Features with Enhanced Resolution}

\author{Nilakshi Senapati}
\email{nilakshisenapati0408@gmail.com}
\affiliation{%
	Department of Physics, Indian Institute of Technology Kanpur, Kanpur 208016, India
}
\author{Abhinandan Bhattacharjee}%
\affiliation{Institute for Photonic Quantum Systems (PhoQS), Paderborn University, Warburger Str. 100, 33098 Paderborn, Germany}
\author{Anand K Jha}
\email{akjha@iitk.ac.in}
\affiliation{%
 Department of Physics, Indian Institute of Technology Kanpur, Kanpur 208016, India
}%

\date{\today}

\begin{abstract}
Imaging through scattering media remains a long-standing challenge in numerous real-world applications, ranging from medical imaging to long-distance sensing. Recently, illumination consisting of a single orbital angular momentum (OAM) mode, which is structured in the azimuthal coordinate, has been shown to provide enhanced resolution for imaging objects with azimuthal features, with the resolution becoming maximum at an optimal OAM value. However, in the presence of scattering, single-mode fields, which are spatially fully coherent, cause the imaging resolution to decrease significantly due to speckle formation.  In this work, we employ azimuthally partially coherent fields and experimentally demonstrate imaging of azimuthal features with enhanced resolution in the presence of scattering. We show that lower degree of azimuthal coherence in such illumination leads to increased robustness against scattering while the azimuthal structure of the illumination ensures enhanced resolution. We derive the condition for best imaging resolution, and we report increase of imaging contrast in scattering from about 7\% to 50\% as the illumination is changed from a single-mode fully coherent field to that of an azimuthal partially coherent field. 
\end{abstract}

\maketitle

Imaging through complex environments such as scattering and turbulent media is important in a wide range of applications, including biological imaging, long-range sensing, and astronomical imaging \cite{katz2014natphot, vorontsov2024photonics, gigan2022jop, bertolotti2022natphy, cheng2025lsa, kogelnik1968josaa}. Over the years, numerous experimental methods have been developed to enhance imaging capabilities through complex media. Among them, several reconstruction-based approaches employ wavefront shaping, phase retrieval, and inverse scattering techniques \cite{park2018alpphot, kim2016laserphotrev, vorontsov2024photonics, bertolotti2022natphy}. However, these methods often rely on prior knowledge or characterization of the complex medium, computationally intensive algorithms, and post-processing. Direct imaging approaches without reconstruction have also been developed using spatially partially coherent fields \cite{redding2012natphoton, wheeler2011appopt, redding2015pnas, bhattacharjee2020pra}. Compared to fully coherent illumination, such fields improve imaging resolution in complex media and are also preferred for information transfer in turbulence \cite{peng2021photonix, liu2013optlett, wang2017appopt, bhattacharjee2020optlet}, digital holography \cite{dubois2004appopt}, and wide-field optical coherence tomography \cite{kim2005jobp, karamata2004optlett}.

Recently, it has been shown that light field structured in the azimuthal coordinate leads to imaging of azimuthal features with super-resolution  \cite{senapati2026apl}. In particular, it was demonstrated that orbital angular momentum (OAM) modes, which are structured in the azimuthal coordinate, leads to super-resolution of objects with azimuthal features, with the resolution becoming maximum at an optimal OAM value for a given azimuthal feature. However, several relevant objects, such as biological specimens \cite{szymborska2013science, loschberger2012cellscience, ruland2021natcomm, lau2012sted, leake2006nature} that have prominent azimuthal features, are found embedded in scattering environments. For such objects, illumination consisting of an OAM mode does not result in super-resolution. This is because an OAM carrying mode is spatially fully coherent, and therefore in the presence of scattering, it causes the imaging resolution to decrease significantly due to speckle formation \cite{goodman2007}. In this Letter, we address this limitation by introducing azimuthally partially coherent illumination, which we realize by generating incoherent mixtures of OAM carrying modes.

\begin{figure*}
	\centering
	\includegraphics[width=1.02\textwidth]{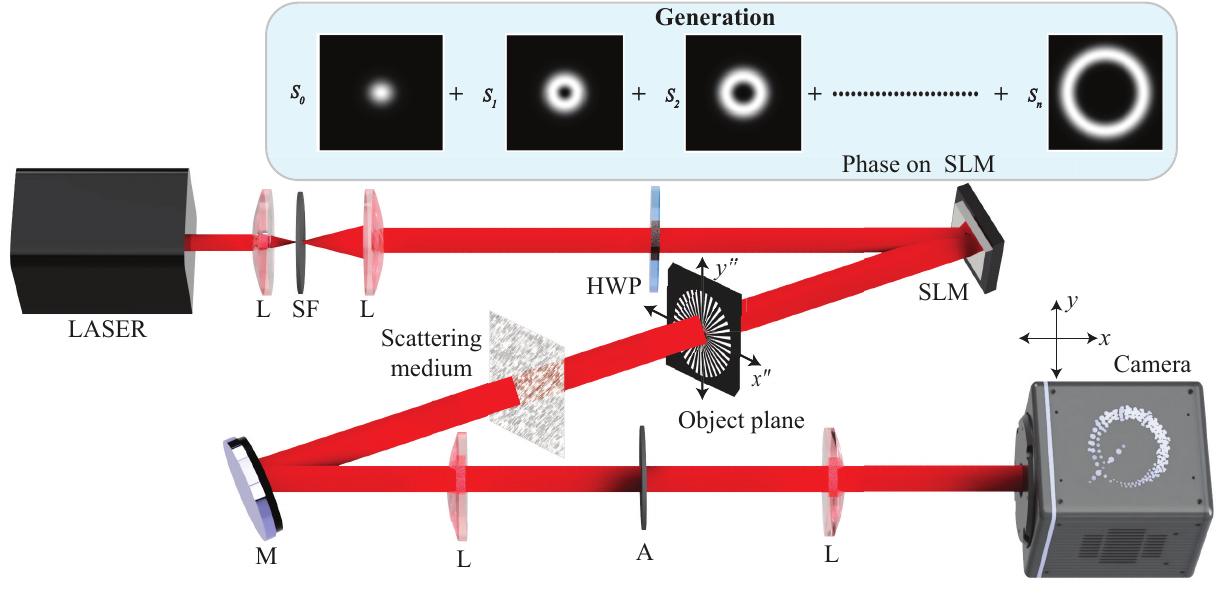}
	\caption{The schematic of the  experimental setup. L: Lens, SF: Spatial Filter, A: Aperture, M: Mirror, HWP: Half Wave Plate, SLM: Spatial Light Modulator. The inset shows the schematic for producing azimuthally partially coherent field as an incoherent mixture of OAM modes.}
	\label{setup}
\end{figure*}

The imaging configuration along with the experimental setup is shown in Fig.~\ref{setup}. The object is placed at $z=0$, the imaging lens at $z=2f$, and the image plane at $z=4f$, where $f$ is the focal length of the lens. The coordinates of the object and the image planes are denoted by $(r'',\theta'')$ and $(r,\theta)$, respectively. The electric field right after the object plane $z=0$ is given by $E_{\rm in}(\theta'')T(\theta'')$, where $E_{\rm in}(\theta'')$ is the electric field of the illumination and $T(\theta'')$ is the object transmission function. Here, we assume that the object possess only azimuthal features, and that the illuminating field is uniform in the radial coordinate. Therefore, the resulting electric field at the image plane can be expressed as  \cite{mertz2019cambridge, goodman2005, senapati2026apl} :
\begin{multline}
	E(r, \theta;{z=4f}) = A e^{-\frac{r^{2}}{\sigma_{p}^{2}}} \iint  T(\theta'') E_{\rm{in}}(\theta'') \exp{\biggl[-\frac{ar''^{2}}{2\sigma_{p}^{2}}\biggr]} \\\times \exp\biggl[-\frac{rr''}{\sigma_{p}^{2}} \cos(\theta - \theta'') \biggr] r'' dr'' d\theta'', \label{E4f}
\end{multline}
where $\sigma_{p}=\frac{2f}{kd}$, $a=\bigl( 1 + \frac{\sigma_{p}^{2}}{2w^{2}} - i\frac{k\sigma_{p}^{2}}{2f} \bigr)$, $k$ is the wave vector, $w$ is the beam size at the object plane, and $d$ is the size of the lens aperture. For conceptual clarity, we consider an azimuthal double-slit object with infinitesimally narrow slit-widths and an angular slit-separation of $2\theta_0$ and thus write the transmission function as $T(\theta'')=\delta(\theta''-\theta_{0}) + \delta(\theta''+\theta_{0})$. The intensity at the image plane can is given by $I(r,\theta;z=4f) \equiv \langle E^*(r, \theta;{z=4f}) E(r, \theta;{z=4f}) \rangle_e$, where $\langle\cdots\rangle_e$ represents the ensemble average over different realizations of the partially coherent illumination. Using Eq.~(\ref{E4f}), we write it as
\begin{align}
	I(r,&\theta; z=4f) \notag \\ = & \left\langle \big{|} E_{\rm{in}}(\theta_0) f(r, \theta - \theta_0) + E_{\rm{in}}(-\theta_0) f(r, \theta + \theta_0)\big{|}^2 \right\rangle_e,	 
\end{align}%
where $f(r,\theta) = A e^{-\frac{r^{2}}{\sigma_{p}^{2}}} \int e^{-\frac{ar''^{2}}{2\sigma_{p}^{2}}} \exp{\bigl[ - \frac{r r''}{\sigma_{p}^{2}} \cos\theta \bigr]} r'' dr''$. The imaging resolution is taken to be maximum when the two slits are imaged with maximum possible contrast, which takes place when the intensity $I(r,\theta=0; z=4f)$ at the center of the two slits ($\theta=0$), is minimum. For calculating the condition for maximum contrast, we write the image plane intensity at $\theta=0$, where the function $f(r, \theta-\theta_0)$ is taken to be symmetric, in the sense that $f(r, -\theta_0)=f(r, \theta_0)=f_0(r)$. Thus
\begin{align}
	I(r,&\theta=0; z=4f) \notag \\ = & |f_{0}(r)|^{2} \bigl[ I_{\rm{in}}(\theta_{0}) + I_{\rm{in}}(-\theta_0) + W(\theta_{0}, -\theta_{0}) + \text{c.c.} \bigr].
\end{align}
Here, $I_{\rm{in}}(\theta_{0}) = \langle E^*_{\rm{in}}(\theta_0) E_{\rm{in}}(\theta_0) \rangle_e$, $I_{\rm{in}}(-\theta_{0}) = \langle E^*_{\rm{in}}(-\theta_0) E_{\rm{in}}(-\theta_0) \rangle_e$, and $W(\theta_{0}, -\theta_{0})=\langle E^*_{\rm{in}}(\theta_0) E_{\rm{in}}(-\theta_0) \rangle_e$ is the cross-spectral density function $W(\theta_1,\theta_2)$  between azimuthal positions $\theta_1$ and $\theta_2$. We consider the illuminating field to be an incoherent mixture of OAM modes, and therefore it can be written as \cite{jha2011pra}:
\begin{equation}
	W(\theta_0, -\theta_0)\rightarrow W(2\theta_0)=\frac{1}{2\pi}\sum_{l=-\infty}^{\infty}  S_{l} e^{i l2\theta_0}, \label{Wdef} 
\end{equation}
where the OAM spectrum $S_{l}$ denotes the weight of the OAM modes. We take $S_l$ to have the Gaussian distribution centred around a mean $l_0$ and having a spectral width $\sigma$, which is assumed to be much smaller compared to $l_0$, that is, $S_l= A \exp\left[-\frac{(l-l_0)^2}{2\sigma^2}\right]$ with $\sum_{l=-\infty}^{\infty} S_l=1$.  Thus,
\begin{align}
	W(2\theta_0)=e^{i l_02\theta_0} \frac{1}{2\pi}  \sum_{l=-\infty}^{\infty}  A \ \  e^{\frac{-l^2}{2\sigma^2}} e^{i l2\theta_0}=e^{i l_02\theta_0} g(2\theta_0), \label{Wdef2} 
\end{align}
where $g(2\theta_0)\equiv A e^{\frac{-l^2}{2\sigma^2}} e^{i l2\theta_0}$ can be referred to as the degree of azimuthal coherence and its width $\theta_{\rm{c}}$ as the azimuthal coherence width \cite{jha2011pra}. We note that $W(2\theta_0)$ and $S_l$ form a Fourier transform pair. This implies that by increasing the width $\sigma$ of the OAM spectrum $S_l$, one can decrease the azimuthal coherence. For the spectrum containing only a single OAM mode, the field remains azimuthally fully coherent, given by $\theta_c=2\pi$ \cite{senapati2026apl}. Using Eqs.~(\ref{Wdef}) and (\ref{Wdef2}), we write Eq.~(\ref{E4f}) as
\begin{align}
	I(r,\theta=0;z=4f) = 2|f_{0}(r)|^{2} \bigl[ 1 + g(2\theta_{0}) \cos(2l_{0}\theta_{0})\bigr]. \label{finalI}
\end{align}
\begin{figure*}[t!]
	\includegraphics[scale=1]{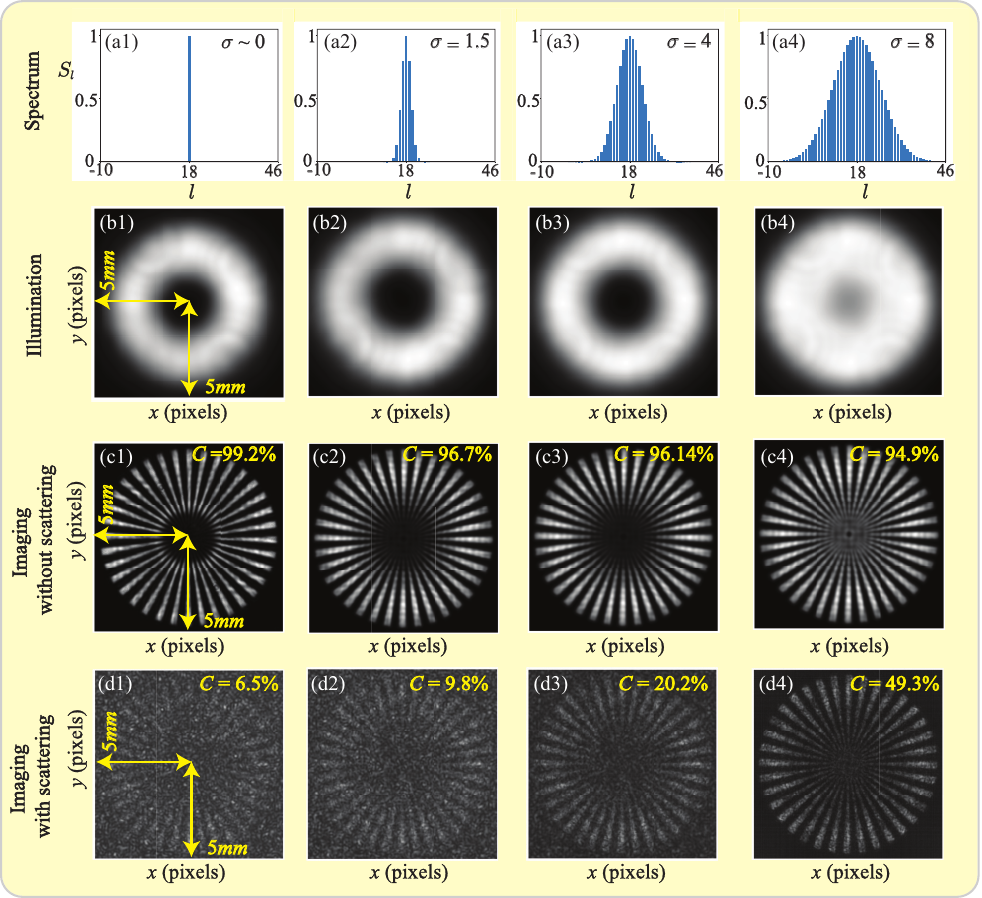}
	\caption{Experimental Results: (a1)-(a4) OAM spectrum of Illumination  with different azimuthal coherence. (b1)-(b4) the corresponding object-plane intensity of the Illuminating fields. (c1)-(c4) Imaging with various illuminating field in the absence of scattering. (d1)-(d4) Imaging with various illuminating field in the presence of scattering.}
	\label{results}
\end{figure*}
As noted above, the maximum imaging contrast is obtained when the intensity $I(r,\theta=0;z=4f)$ achieves its minimum. It can be shown that the conditions for a minimum, that is, $\frac{dI}{dl_{0}}=0$ and $\frac{d^{2}I}{dl_{0}^{2}}>0$, are satisfied when
\begin{equation}
	l_{0} = \frac{(2n+1)\pi}{2\theta_{0}}=l_{\rm opt}, \label{lopt}
\end{equation}
where $n=0, \pm 1, \pm 2, \pm 3 \dots$. We note that since $g(2\theta_{0}) \neq 1$ for an azimuthally partially coherent field, the imaging contrast can never exceed that with azimuthally fully coherent field in which case $g(2\theta_{0})=1$. Nonetheless, condition for maximum possible contrast remains the same for an azimuthally partially partially coherent field as that in the case of a azimuthally fully coherent field \cite{senapati2026apl}. This is due to the fact that when imaging with illumination that are structured in the azimuthal coordinate, the condition for optimum structuring of the illumination is dictated by the azimuthal features of the object. Therefore, in the absence of a scattering medium, azimuthally structured fully coherent fields gives the best possible imaging contrast \cite{senapati2026apl}. However, in the presence of a scattering medium, the imaging contrast decreases due to speckle formation. Next, we show how the contrast lost due to scattering can be recovered by decreasing the azimuthal coherence of the illuminating field.

The experimental setup is shown in Fig.~\ref{setup}. A 5 mW He--Ne laser beam is incident on a Holoeye Pluto spatial light modulator (SLM), on which phase patterns are displayed to generate azimuthally partially coherent field using the Arrizón method \cite{arrizon2007josaa}. The Gaussian OAM spectrum with mean $l_{\rm opt}$ and width $\sigma$ is generated by sequentially displaying onto the SLM the phase patterns corresponding to various OAM modes \cite{bhattacharjee2019jopt}. The modal weights $S_l$ are controlled by choosing the display time of each phase pattern proportionately, and the total acquisition time of the camera is set sufficiently long to record the average intensity over all the displayed modes. The object is imaged onto the camera using a $4f$ imaging system consisting of two lenses with focal lengths $f=400$ mm. A Thorlabs R1L1S2N resolution target is used as the object. For this object, $\theta_0=\pi/18$, yielding $l_{\rm opt}=18$.  The scattering medium is placed between the object and the first lens and consists of a Thorlabs DG600 diffuser together with a single layer of lens tissue. Here, the scattering strength ($\alpha$) \cite{bhattacharjee2020pra} is defined as the ratio of the intensities in the presence and absence of the scattering medium.  For the scattering medium used in our experiment, we measure $\alpha=0.61$. Figure~\ref{results} shows the experimental results with azimuthally partially coherent fields having $l_{\rm opt}=18$. The first column of results are with azimuthally fully coherent mode $l_{\rm opt}=18$, indicated by $\sigma\approx 0$. The subsequent three columns of results are with three different azimuthal partial coherence, namely, $\sigma=1.5$, $\sigma=4$, and $\sigma=8$.

Figures~\ref{results} (a1)-(a4) show the spectrum for the four different fields while Figs.~\ref{results} (b1)-(b4) show the corresponding illumination. Figures~\ref{results} (c1)-(c4) present the imaging results in the absence of scattering while Figs.~\ref{results} (d1)-(d4) present the corresponding images in the presence of scattering. We find that as the azimuthal coherence is reduced, the scattering-induced degradation is progressively overcome and the image quality improves. To quantify this improvement, we calculate the image contrast, $C=\frac{I_{\rm{max}}-I_{\rm{min}}}{I_{\rm{max}}+I_{\rm{min}}}$, where $I_{\rm{max}}$ and $I_{\rm{min}}$ are the the minimum and maximum intensities at the image plane. We note that the illumination has a central dark region because of the absence of modes closer to $l=0$, and therefore, for calculating the imaging contrast, we exclude this central dark region up to a certain radius. First of all, we find that in the absence of scattering all the four fields are able to image the object with close to perfect contrast, even though the imaging contrast for the azimuthally fully coherent field is maximum. However, in the presence of scattering the azimuthally fully coherent illumination is not able to image the object as the contrast is only 6.5\% [Fig.~\ref{results} (d1)]. However, as $\sigma$ is increased the imaging contrast increases up to 49.3\%. Thus by reducing the azimuthal coherence of the illuminating field the scattering-robust imaging of azimuthal features can be achieved.

In summary, in this letter, we have experimentally demonstrated  imaging of azimuthal features with enhanced resolution in the presence of scattering, without employing any image-reconstruction techniques. Since such illumination is structured in the azimuthal coordinate, it images azimuthal features with enhanced resolution. At the same time, the lower degree of azimuthal coherence leads to increased robustness against scattering. Our imaging approach can have important implications for biological specimens with well-defined azimuthal features, such as nuclear pore complexes, centrioles, bacterial flagellar motors through scattering tissue \cite{szymborska2013science, loschberger2012cellscience, ruland2021natcomm, lau2012sted, leake2006nature}.

\paragraph*{Conflict of Interest:} The authors have no conflict to disclose.

\paragraph*{Data Availability:} Data underlying the results presented in this paper are not publicly available at this time. The data that support the findings of this study are available on request from the corresponding authors.

\paragraph*{Acknowledgements:} N.S. thanks the Prime Minister’s Research Fellowship (PMRF), Government of India, for financial support. We acknowledge financial support from the Science and Engineering Research Board through grants STR/2021/000035 and CRG/2022/003070, and from the Department of Science \& Technology, Government of India through the QuEST grant DST/ICPS/QuST/Theme-I/2019 and  the National Quantum Mission (NQM) grant DST/FFT/NQM/QSM/2024/3 for quantum imaging.

\nocite{*}
\bibliography{ref}

\end{document}